\documentclass[12pt]{article}
\usepackage{graphicx,amsmath}
\usepackage{color}

\newlength{\dinwidth}
\newlength{\dinmargin}
\def\be{\begin{equation}}   
\def\ee{\end{equation}}  
\def\bea{\begin{eqnarray}}                      
\def\eea{\end{eqnarray}}
\def\ch1{$\chi(1^+)$}
\def\lapproxeq{\lower .7ex\hbox{$\;\stackrel{\textstyle                                                    
<}{\sim}\;$}}                                                    
\def\gapproxeq{\lower .7ex\hbox{$\;\stackrel{\textstyle                                                    
>}{\sim}\;$}}

\begin{document}

\begin{flushright}                                                    
IPPP/17/14 \\
\today \\                                                    
\end{flushright} 

\vspace*{0.5cm}

\vspace*{0.5cm}
\begin{center}
{\Large \bf Can invisible objects be `seen'  
via forward proton} \\ 
\vspace{0.2cm}
{\Large \bf  detectors at the LHC?}\\
\vspace*{1cm}

V.A. Khoze$^{a,b}$, A.D. Martin$^{a}$ and M.G. Ryskin$^{a,b}$\\

\vspace*{0.5cm}
$^a$ Institute for Particle Physics Phenomenology, Durham University, 
Durham, DH1 3LE, UK\\ 

$^b$ Petersburg Nuclear Physics Institute, NRC `Kurchatov Institute', 
Gatchina, St.~Petersburg, 188300, Russia \\
\end{center}

\begin{abstract}

We discuss the possibility of identifying invisible objects (e.g. dark matter) via  missing mass in Central Exclusive Processes; that is, in events where {\it only} forward protons are detected at the LHC.  We show that the signal must have a cross section greater than 0.25 fb in order that it can be detected.  We estimate the huge background 
caused by soft proton dissociation and evaluate the requirements of the detector `veto' system needed to sufficiently suppress these events.   In addition, we discuss the related process leading to the possible identification of the particles from the so-called `compressed-mass Beyond the Standard Model (BSM)' scenarios.
In this case the background is not so severe.

\end{abstract}
\vspace*{0.5cm}

\section{Introduction} 
The existence of dark matter (DM) in the Universe
 is commonly accepted as the explanation of many experimental phenomena in astrophysics and cosmology (for a recent review see \cite{Freese:2017idy}).
Within the standard model of cosmology 
the data
collected by the Planck collaboration \cite{Ade:2015xua}
 imply that DM constitutes around 80$\%$ 
of the total matter content of the Universe. 
The new DM
objects should  be
massive, should not carry an electric charge and could weakly interact with the
common matter. DM particles are predicted by numerous BSM  models, such as,
for instance the $R$-parity conserving supersymmetry (SUSY).
A review of these models can be found in \cite{Abercrombie:2015wmb,Boveia:2016mrp}.
If  non-gravitational interactions between DM and SM particles exist
these hypothetical particles could be produced at hadron colliders.
Despite an intensive experimental efforts the dark matter searches so far have
been unsuccessful. Not surprisingly,
the search for DM is one of
the main objectives of the LHC physics program since the DM particles
could appear in proton-proton $(pp)$ collisions. DM particles on their own
would not leave a detectable signal in the LHC detectors, but when produced in association
with a visible SM particle  
$X$ (which may be a $g, q, \gamma, Z, W$ or $h$)
they would provide event topologies with a transverse momentum {\it imbalance}.
A large missing transverse energy $E_T$ will be then observable in the detector.

So far no signs of
Dark Matter particles have been observed in particle physics
experiments, and currently at the LHC only new
constraints on the parameter
spaces of the theoretical models
have been set\footnote{See, for example, \cite{Kahlhoefer:2017dnp}
for a review of recent  searches for dark matter at the LHC.}.
 
 However, in some scenarios the missing $E_T$ may be not large. In particular, 
this may be the case when the invisible object is produced via photon-photon fusion, see, for example \cite{lowET}. 
Here we discuss the possibility of searching for such invisible objects in exclusive events
\be
pp~\to p+{\rm invisible}+p,
\ee
where the missing $E_T$ is relatively small, by studying the missing-mass distribution 
 with respect to the two outgoing forward protons. In other words, to search for exclusive events with missing longitudinal energy.
The serious problem is that the LHC detectors are not `hermetic'. As a consequence, there is a large probability that the events have secondaries with large rapidities that escape detection and so carry away the corresponding longitudinal energy.

\subsection{Cross section and background}
Since the missing-mass signal is obtained by observing only the forward protons, and not 
observing the vertex in the central detector, we have no possibility of excluding a `background' pile-up event. In particular, in the same bunch crossing we may detect a proton coming from a single diffractive 'right' beam  dissociation and another leading proton from another event with a 'left' beam dissociation. The cross section of low-mass dissociation  measured
by TOTEM at $\sqrt s=7$ TeV is $\sigma^{\rm D}=2.6\pm 2.2$ mb~\cite{TOTEM-D}.
That is, the probability of one side dissociation is of the order of
\be
\frac{\sigma^{\rm D}}{2\sigma_{\rm tot}}~\sim~ \frac{2.6~{\rm mb}}{2\times 100~{\rm mb}}~\sim ~1\% .
\ee 
At high LHC luminosity, when the mean number of interactions in each bunch crossing is about $\mu=50$ we will observe such a `wrong' configuration (two leading protons from two different events) in typically each fourth bunch crossing.
For this reason it looks better to work at lower luminosities, when $\mu\sim 1$, using, for example, the ALICE detector. 

Anyway, we have no chance to observe more than one `invisible' event per bunch crossing.  This means that the lowest `invisible' cross section that we can discuss is of the order of 
\be
\label{sig}
\sigma_{\rm min}~\sim ~\sigma_{\rm tot}\left(\frac{\Delta t}{T_{\rm  run}}\right)~\sim~ 100~\mbox{mb}\left(\frac{ 25~ \mbox{ns}}{10^7\ \mbox{s}}\right)~\sim ~0.25\ \mbox{fb}\ , 
\ee
here $\Delta t$ is the bunch-bunch spacing and $T_{\rm run}$ is the time of the LHC run.  For $\mu\sim 1$ the ratio $T_{\rm run}/\Delta t$ is the total number of $pp$ collisions.
  
Recall that the cross section of low-mass dissociation is of the order of 1 mb (where we have a leading proton with the momentum fraction $0.7<x_L<0.95$, but no secondaries in the central detector (or calorimeters)).  The corresponding cross section of double low-mass dissociation (both protons) is about 0.1 - 0.2 mb (see e.g.~\cite{elastic}). This is in accordance with the factorization relation
\be
\label{eq:fact}
\sigma^{\rm DD}=\frac{\sigma^{\rm SD}\cdot\sigma^{\rm SD}}{\sigma_{\rm el}}\ .
\ee
Inserting $\sigma_{\rm el}=25$ mb and $\sigma^{\rm SD}=1$ mb in (\ref{eq:fact}), we would expect
 $\sigma^{\rm DD}=0.04$ mb. However it is known that the exact factorization relation is violated by the different $t$-behaviour of the cross sections and the different gap survival probabilities (see, for example,~\cite{elastic} for details). In particular,  at 7 TeV in the mass interval $M_X=3.4-8$ GeV, the ratio is observed to be  $\sigma_{\rm DD}\sigma_{\rm el}/\sigma_{\rm SD}^2=3.6$~\cite{TOT-fac} and not 1.
That is, we expect $\sigma^{\rm DD}$ to be about 4 times larger, $\sigma^{\rm DD}\sim 0.16$ mb, than that given by (\ref{eq:fact}).
To reach an acceptable signal-to-background ratio, $S/B\sim 1$, we therefore must suppress such dissociation of both the left and right outgoing protons by about 6 orders of magnitude. Assuming a factor  4$-$10 violation of the factorization prediction we conclude that in order to have an `acceptable' background dissociation, the value of  $\sigma^{\rm SD}$ should be reduced down to a few nb.  \footnote{Even with  $\mu=1$ there is a probability of having  
two events in one bunch crossing and to produce two leading protons from two different events, but this probability will be less than the probability of the event coming from `double dissociation'.}

\subsection{Outline}

In this paper we estimate whether it is possible that LHC detectors can obtain the necessary suppression of the background to identify an invisible signal of cross section of the order of 0.25 fb, see (\ref{sig}).

The major channels of low-mass proton dissociation are the formation of
the nucleon  resonances, like $N^*(1440)$ (with the same spin, parity as the proton), and the group of resonances in 1700 MeV mass region (see e.g. sect.6.2 of~\cite{albery}), with dominant decays into the $N\pi$ and $N\pi\pi$ states. 
Section~\ref{sec:s2} is devoted to the background arising from the decays $p\to p+$ charged neutral final states. In Section~\ref{sec:22} we consider the possibility of the Zero Degree Calorimeter (ZDC) (for example~\cite{ZDC} for ALICE) observing the $\pi^0$ from $N^*\to p+\pi^0$ decay.   In addition, in Section~\ref{sec:21} we discuss photon bremsstrahlung ($p\to p+\gamma$), which also may produce a leading proton with not too small cross section (see~\cite{brem}), and in Section~\ref{sec:23} we study the radiative
$N^*\to p+\gamma$ decays which have sizeable branching ratios (in particular,  Br$(N(1440)\to p\gamma)\sim 0.04$\%, Br$(N(1535)\to p\gamma)\sim 0.2-0.3$\% ~\cite{PDG}). Finally, in Section~\ref{sec:24}   we evaluate the contribution of the so-called Drell-Hiida-Deck diagram~\cite{DHD}, but this contribution is much smaller. In Section~\ref{sec:s3} we estimate the chance of vetoing the charged pions from $N^*\to p+\pi^+\pi^-$ decay with Forward Shower Counters (FSC) \cite{FSC} (and ADA and ADC of ALICE~\cite{ADA}). 

We do not consider here other background processes (like Central Exclusive Production of ZZ-boson pairs with their subsequent decay to four  neutrinos, or other SM reactions with neutrinos and charged leptons or photons which for some 
reason were not observed in  central detector). These processes have a much smaller cross section (see e.g.~\cite{arman}) than those for proton dissociation.  There could be another source of background caused by beam halo particles. It depends on the details of the beam optics and we will not consider it.

In Section~\ref{sec:s4}  we consider the possibility of identifying other particles by missing mass signals. We discuss the `compressed mass BSM' scenario. In this case the process is not completely invisible but contains relatively low $E_T$ jets or leptons which will be difficult to distinguish from the SM  background in usual inclusive events. 

All the numerical estimates are done for the initial proton-proton energy $\sqrt s=13$ TeV assuming that the leading protons with the momentum fraction, $x_L$ are detected within the interval $0.82<x_L<0.96$~\footnote{We thank Kenneth Osterberg for useful discussions concerning the `safe' values of $x_L^{\rm min}$ and $x_L^{\rm max}$ for the Roman pot forward detectors at the LHC.}.

There is a chance that a larger $x_L$ interval may be covered at the LHC using the Beam Line Monitor (BLM) system as the forward proton detector \cite{BLM}.
This method is
complimentary to the present experimental installations
for forward proton detection (TOTEM, ALPHA, CT-PPS, AFP). It
would allow the experiment to greatly increase the missing mass coverage in Central Exclusive Processes.  Recall, that the conventional detection system, based on the Roman Pot
detectors, is limited by the locations of the pots and the allowed transversal
approach to the beam, see \cite{FP420}. 
On other hand, in the central production process, $pp\to p + X + p$, a final state proton, which exits the LHC beam vacuum chamber at
locations determined by its fractional momentum loss, may be detected by the BLM system. In this way its momentum will be measured. That is, the BLM system will act as a forward proton detector
~\cite{BLM}. However, at the moment, we consider the acceptance of the present Roman Pot detectors.

\section{Background from proton to neutral(s) final states   \label{sec:s2}}

In this section we consider the background to the invisible signal coming from the forward proton decays $p\to p+\gamma,~N^*\to p+\gamma$ and $~N^*\to p+\pi^o$.  These are, respectively, the subjects of subsections \ref{sec:21}, \ref{sec:23} and \ref{sec:22} below. Such backgrounds may be rejected by an appropriate Zero Degree detector.

\subsection{Bremsstrahlung: $  p\to p+\gamma$  \label{sec:21}}
We start from the simplest process of photon bremsstrahlung of the proton.
The probability to radiate a photon during elastic $pp$ scattering is suppressed by the QED coupling $\alpha^{\rm QED}/\pi$. However, in comparison with a 1 fb cross section from a possible BSM signal, or from the cross section of single dissociation $\sigma^{\rm SD}\sim 1$ nb allowed for the background processes of interest, the bremsstrahlung background is quite large. Recall that it is proportional the elastic $pp$ cross section of 25 mb.

To make a numerical estimate we use the formula from~\cite{brem} supplemented by the nucleon form factor 
\be
F_N=\left(1+\frac{m_p^2-p^2}{0.71~{\rm GeV}^2}\right)^{-2}
\ee
 at each $\gamma p$ vertex, which accounts for the off-shellness ($p^2\ne m_p^2$) of the intermediate proton.  Here we use
the usual dipole form of this form factor. A more detail discussion of the $pp\to (p\gamma)+p$ reaction can be found in~\cite{sz-brem}, where it was shown that some other mechanisms of dissociation into $ p\gamma$ give  smaller cross sections and are less important.

\begin{figure} [h]
\begin{center}
\vspace{-5.0cm}
\includegraphics[clip=true,trim=0.0cm 0.0cm 0.0cm 0.0cm,width=13.0cm]{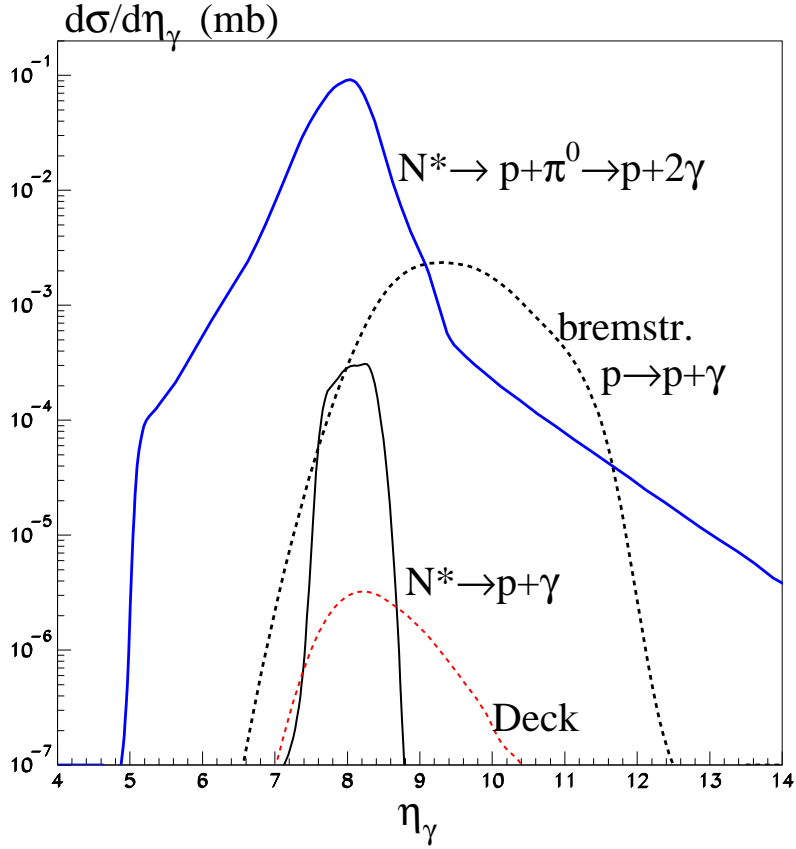}
\caption{\sf The background to the exclusive invisible signal from the dissociation of the forward going protons to proton plus  neutral final states. The curve for $N^*\to p\gamma$ decay is calculated assuming the $\langle M_X \rangle$=1.7 GeV, while for the $N^*\to p\pi^0$ decay $\langle M_X \rangle$=1.9 GeV was used. The results do not depend significantly on the value of $M_X$. Here, it is assumed that $E_\gamma>100$ GeV.}
\label{fig:f1}
\end{center}
\end{figure}

As seen from the dashed line in Fig.~\ref{fig:f1} there is rather large bremsstrahlung at relatively large angles (more than 0.3 mrad)~\footnote{Unfortunately due to the relatively large mass of the proton the photon is emitted at comparatively large angles.}. That is, in the rapidity region ($\eta<8.8$)  not fully covered by the present ZDCs. From this undetected region we collect up to 1 $\mu$b. To suppress this cross section down to 1 nb an enlargement of the ZDC coverage by a factor of about 4 will be needed; that is, up to 1.2 mrad (corresponding to $\eta\simeq 7$).  

\subsection{$N^*\to p+\pi^0$ dissociation   \label{sec:22}}
An even stronger neutral background arises from $\pi^0$ production. Here there is no suppression due to the small QED coupling. Moreover, after the subsequent $\pi^0\to \gamma\gamma$ decay the final photons may be  emitted at relatively large angles
(see the heavy solid curve in Fig.~\ref{fig:f1} which was calculated for $\langle M_X \rangle$=1.9 GeV)~\footnote{The $p\to p+\pi^0$ dissociation was calculated in~\cite{sz-pi0}. However, in this paper the kinematics of the final $\pi^0\to  \gamma\gamma$ decay was not considered.}. On the other hand, to exclude such events it may be sufficient to observe only one of the two high energy photons.
That is, first, we have to be sure that the probability for both photons to miss the ZDC detector is negligibly small.

The production and decay of  $N^*$ resonances in the 1440$-$1900 mass region may be responsible for about 50\%  of the low-mass proton dissociation. The probability of their decay to $p+\pi^0$ channel is about 20$-$25\%. So, this channel has a cross section $\sigma^{\rm SD}(p\to p+\pi^0)\sim 0.1$ mb.

We assume the simplified case where the angular and momentum distributions are driven only by the phase space corresponding to the $N^*\to p\pi^0$ decay.  In addition to the angular distribution of the final photons generated by the $\pi^0\to \gamma\gamma$ decay kinematics, we have to account for the non-zero momentum transferred in the initial $p\to N^*$ dissociation.
At present there are no LHC data on the $t$ distribution in low-mass dissociation. The only preliminary experimental  observation is the $t$-slope $B_D=10.1$ GeV$^{-2}$ measured by TOTEM~\cite{B-10} for the $M_X=3.4-8$ GeV mass interval.
So, here we  assume a slope $B_D=10$ GeV$^{-2}$. Next, we assume that the  leading protons are detected within the $x=0.82-0.96$ interval.

The expected pseudorapidity distribution of the final photons is shown in Fig.1.
The worst situation is when both photons from the $\pi^0$ decay are emitted at angles larger than that covered by ZDC. To avoid such a possibility the calorimeter must cover at least the $\pm 1.5$ mrad interval; then the cross section not vetoed by the ZDC is less than 1 nb  for any   $M_X<2.2$ GeV.  For a smaller coverage of, say, $\pm$1.2 mrad we have, at a mean $M_X$ of 2.2 GeV, a cross section up to 0.3 $\mu$b for events where both photons are missed by the ZDC. 


Next there is about 3\% probability that only one photon will be missed and that the other will go into the angular interval ($\theta<1.5$ mrad) covered by ZDC. (It will reduce to 1.4\% probability in the case of a larger $\pm$2 mrad coverage when there is more chance to detect both photons.) A computation shows that this 3\% corresponds to a cross section of $\sigma^{\rm SD}$=100 nb. To suppress the background down to 1 nb we  need factor 10$^2$. That is, this photon must be observed with the efficiency better than 99\%; in other words, the calorimeter should have more than 5 radiation lengths.

Note that with the $\pm$ 1.5 mrad coverage the probability to observe both photons from $\pi^0$ decay in the ZDC is significant (more than 10\%). So, it will be possible to check experimentally the contribution of each particular nucleon resonance.

\subsection{Resonance decay $N^*\to p+\gamma$   \label{sec:23}}
Recall that the radiative
$N^*\to p+\gamma$ decays  have non-negligible branchings. In particular,  Br$(N(1440)\to p\gamma)\sim 0.04$\%, Br$(N(1535)\to p\gamma)\sim 0.2-0.3$\%~\cite{PDG}. 
 That is, we start from the cross sections of about 1 $\mu$b.
For the $N^*\to p+\gamma$ decays the probability to miss the photon
(with the $\pm$ 1.5 mrad coverage of the ZDC) is practically negligible for $M_X<2.2$ GeV. For larger masses the probability of two-particle decay decreases. 
Thus, even with some probability to miss a photon (e.g. the probability is about 1.5$\%$ for $M_X=2.5$ GeV) the background from the radiative decay of a `heavy' nucleon resonance, $N^*\to p\gamma$, is not a problem. 
 On the other hand, multiparticle decay modes which contain additional secondaries will be easier to veto by observing at least one of the extra particles. However, we still need a good  efficiency to detect a single energetic photon in the ZDC. If the mean $M_X=1.7$ GeV then we require a suppression of about 200. Thus, again, the calorimeter should have more than 5-5.5 radiation lengths.

\subsection{Deck diagram   \label{sec:24}}
Another possible background arises from the Drell-Hiida-Deck diagram~\cite{DHD}, in which the emission of a $\pi^0$ from a proton is followed by elastic $\pi^0+p_{\rm target}$ scattering. However, the cross section is small in the $0.82<x_L<0.96$ interval. If we assume 
$\sigma(\pi^0+p)_{\rm tot}=60$ mb, an elastic $\pi+p$ slope of $B=10$ GeV$^{-2}$
and $g^2_{\pi NN}/4\pi=13.75$, then the cross section is about 1 nb, which is negligible.

\section{Background from $p\to p\pi^+\pi^-$  \label{sec:s3}}

Now we are concerned with the detection of forward going charged pions.
As before we assume the unpolarized decay which is driven just by the phase volume
available for the decay,  and that the $t$-slope of proton dissociation is $B_D=10$ GeV$^{-2}$. We calculate the probability to observe one or both charged pions within the forward {\it rapidity} interval in the events with the momentum fraction of both leading protons in the interval $0.82<x_L<0.96$.
Moreover, motivated by the triple-Regge analysis of~\cite{KKPT}, we assume a mass distribution of the form 
\be
M^2_X\frac{d\sigma}{dM^2_X}~\propto ~1+\frac{2~{\rm GeV}}{M_X}.
\ee
given by the triple-Regge PPP and PPR terms (where P and R denote a pomeron, and a secondary reggeon respectively).

\begin{figure} [h]
\begin{center}
\vspace{-5.0cm}
\includegraphics[clip=true,trim=0.0cm 0.0cm 0.0cm 0.0cm,width=12.0cm]{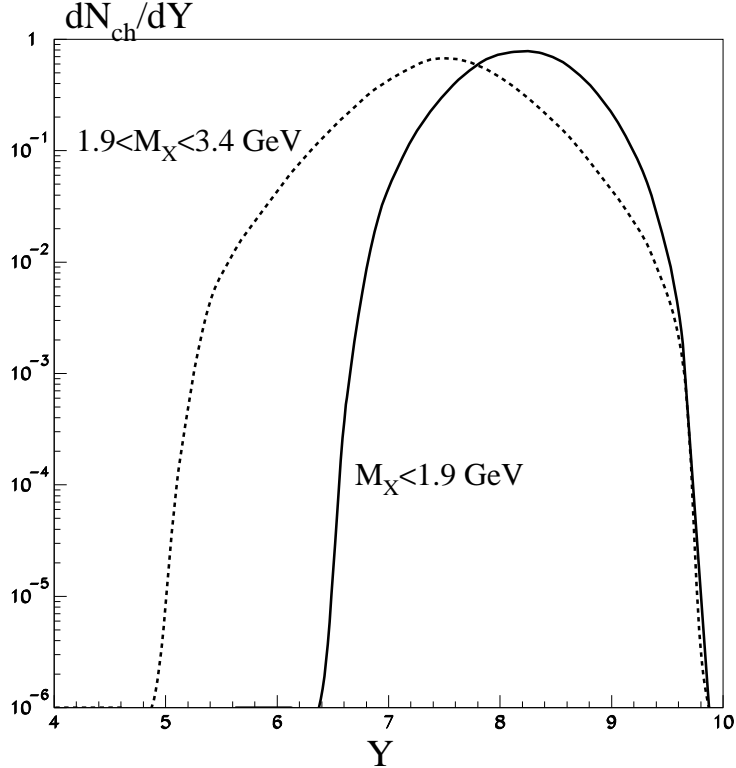}
\caption{\sf The background to the exclusive invisible signal from the dissociation of the forward going protons to charged pions: $p\to p\pi^+\pi^-$. $M_X$ is the mass of the dissociating system. The rapidity distribution is normalized to $\int(dN_{\rm ch}/dY)dY=1$.}
\label{fig:f2}
\end{center}
\end{figure}

As seen from Fig.~\ref{fig:f2}, the `veto' detector must cover the {\it rapidity} interval from 5.5 to 9.5.
Next, it is important to have extremely low cuts (in the 10 MeV range) on the minimum pion transverse momentum, $p_T$. 
Already with the cut $p_T>100$ MeV we have up to 0.5$\%$ (inadmissible) probability to miss both pions from the decay of a $M_X<1.9$ GeV system.
Without a $p_{T{\rm min}}$ cut, the probability that both pions will have the rapidity outside the $5.5<Y<9.5$ interval is negligible (less than $10^{-6}$). On the other hand the probability to miss one
 pion from the decay of a $1.9<M_X<3.4$ GeV system observing 
the $5.5<Y<9.5$ rapidity interval is small, about 1.2$\times 10^{-4}$. That is, to reach the required cross section we have to suppress this background 100 times. This should be possible; $\sim 5$ radiation length is needed. The problem, however, is that here we are dealing with charged particles whose trajectories are affected by the magnetic field. Thus a detailed detector simulation is needed to confirm the efficiency of the `veto'.   


\section{Compressed mass spectra scenarios  \label{sec:s4}}

  Nowadays a lot of attention is paid to 
 BSM models where new particles
are nearly degenerate in mass. Such a mass spectrum is often referred to as
`compressed'. Compressed spectra can arise for example in natural 
SUSY, see for instance
\cite{Dimopoulos:1981zb,Witten:1981nf,Dine:1981za,Sakai:1981gr,Dimopoulos:1981au,Wang}.\footnote{For a discussion of simplified models of dark matter with small mass splitting between dark matter and its visible partner see, for example, \cite{vvk} }
The  present lack of evidence for SUSY at the LHC
makes it attractive to study the compressed scenario
when, in a LHC collision, the lightest supersymmetric particle (LSP),
often assumed to be neutralino, 
 is close in mass to the parent sparticle 
(e.g. chargino).  The neutralino in turn would provide
a viable DM candidate.
For instance, as pointed out in e.g. \cite{Wang,Baer:2014kya,Han:2014kaa,
Baer:2016new}, light higgsinos
could have a compressed mass spectrum resulting in an experimental
signature with comparatively soft leptons (or jets) and missing energy.
If the momentum of a visible SM probe (e.g. charged lepton) in the decay
of the second-to-lightest object is sufficiently small,
the compressed-mass class of theoretical models becomes
very challenging for typical
SUSY searches at the LHC,  because of the trigger requirements
and the difficulty to distinguish experimentally
the signal from the SM backgrounds.
\begin{figure} [h]
\begin{center}
\vspace{-.5cm}
\includegraphics[clip=true,trim=0.0cm 0.0cm 0.0cm 0.0cm,width=18.0cm]{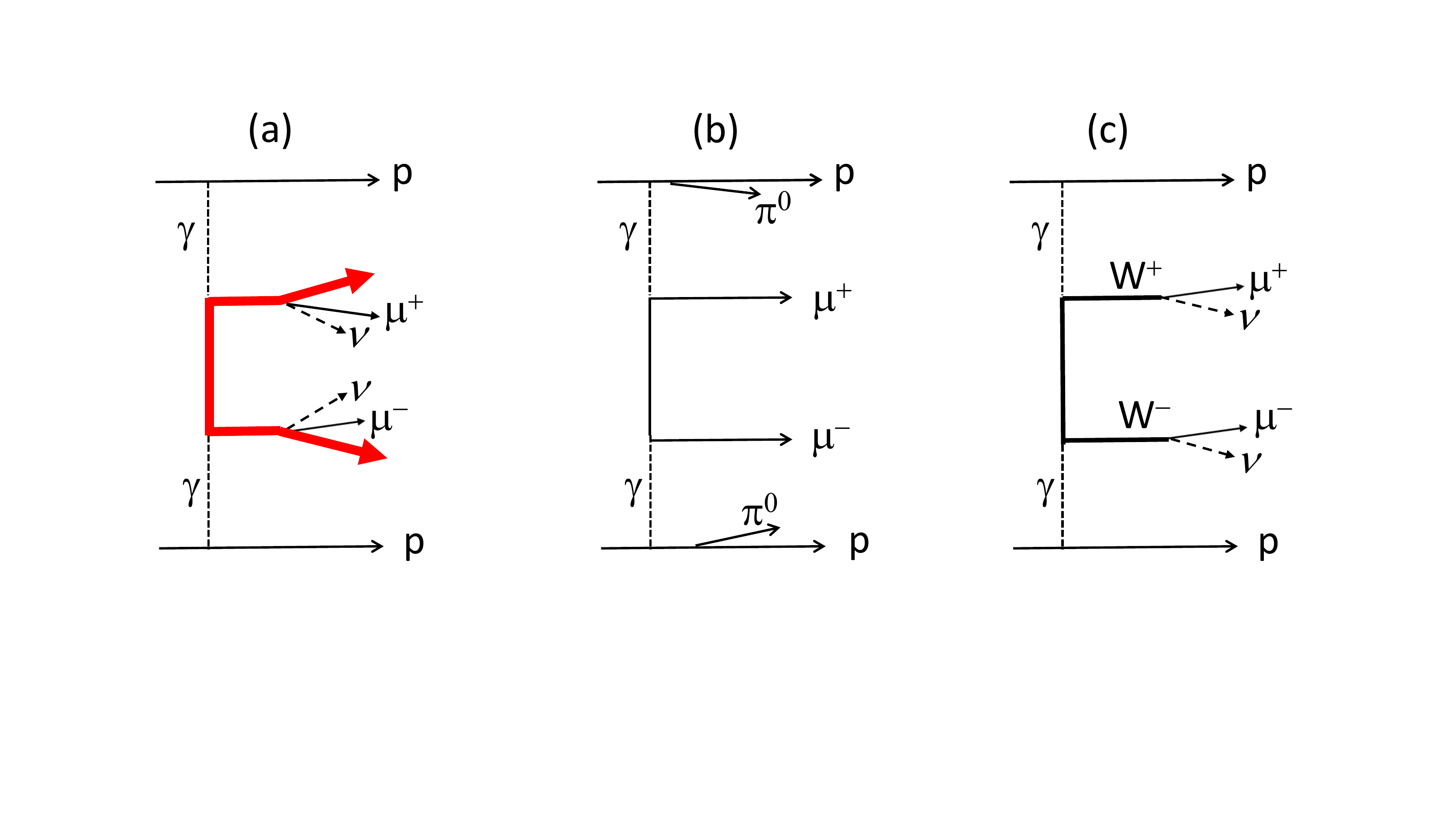}
\vspace{-4.0cm}
\caption{\sf (a) The signal, $pp\to p+\mu^+\mu^- +p$, for a heavy invisible particle, shown by the thick (red) line, in a BSM compressed scenario.  (b,c) The background contributions coming from proton dissociation in QED lepton-pair production and from $W^+W^-$ production respectively. }
\label{fig:f3}
\end{center}
\end{figure}
Therefore, here,  we study a special situation where the new secondaries produced in the event are not completely invisible, but decay into a heavy invisible particle plus a relatively `soft' lepton or minijet. As an example it may be a chargino-pair produced via the photon-photon fusion, in which both charginos decay into the invisible LSP, and a $\nu$, plus  a charged lepton detected with $p_T\sim 3-10$ GeV, see Fig.~\ref{fig:f3}(a). In such a case the missing transverse energy is rather low, while the LSP-pair carries away large longitudinal energy.

Unfortunately again we have to work at  `low' luminosity with $\mu\sim 1$. The problem is that even with  good timing, when we can identify the vertex in central detector which corresponds to the observed leading protons, we cannot be sure that we know if the energy has been carried away by the $\pi^0$ -- the calorimeter is unable to identify the vertex of emission with sufficient accuracy. 

Next, we have to check that this was the prompt leading proton 
and not a proton from $N^*$ decay. That is we need a veto for the production of high rapidity pions or photons.
 There are two problems. First, we have to suppress the SM background. Second, in order to know the missing longitudinal energy caused by some BSM process, we have to be sure that this fraction of the proton momentum was not carried away by an unobserved
pion, photon or other SM secondary.

 However, for the compressed scenario,  the required suppression 
of the background is not so strong. Indeed, the main backgrounds are
(i) the QED production of a lepton (or minijet) pair,
(ii) production of a $WW$-pair with the decay to  low $p_T$ leptons, see Figs.~\ref{fig:f3}(b) and (c) respectively. Finally there may be Central Exclusive Production (CEP) of the pair of $D\bar D$ (or $B\bar B$) mesons following by the decays, like $D^+\to \pi^0e^+\nu$, with the low energy neutral pion  not observed in the central detector. 

The expected exclusive cross sections are about 6 (0.7) pb for the pair of muons with  $p_T>5\ (10)$ GeV and about 70 fb for $WW$-pair production~\cite{el-difr}. The exclusive $D\bar D$ cross section may be estimated based on the formulae for the charm quark pair production from sect.3 of~\cite{Prosp}. It is of the order of 1 nb.

The cross section of analogous processes with the proton dissociation can be roughly evaluated based on the factorization relations. 
\be 
\frac{\sigma_{\rm hard}(p\to N^*)}{\sigma_{\rm hard}(p\to p)}=\frac{\sigma_{\rm SD}(p\to N^*)}{\sigma_{\rm el}(p\to p)}
\label{fact1}\ .
\ee
In hard processes, here, we assume the same probability of dissociation as that in soft (quasi elastic) interactions.

Strictly speaking such factorization relations are violated due to the different $t$-dependences and gap survival probabilities (see e.g.~\cite{el-difr} for details). In particular,  at 7 TeV the ratio $\sigma_{\rm DD}\sigma_{\rm el}/\sigma_{\rm SD}^2=3.6$~\cite{TOT-fac,B-10} and not 1.
On the other hand, the violation of (\ref{eq:fact}) is not large (less than factor of 10). So, the possibility of proton dissociation does not enlarge the final cross section more than a few times. 
Thus to reduce the background down to 1 fb
we have to suppress the $WW$ background about 100 times. Recall that we select events with a relatively low $p_T\sim 10$ GeV, while the typical $p_T$ in W decay is about $M_W/2\sim 40$ GeV. The probability to have so low $p_T$ for {\em both} W decays is of about $(p^2_T/(M_W/2)^2)^2\sim 1/4^4=0.4$\%. That is the expected $WW$ background is less than 1 fb, without any additional selection.

In the case of a background caused by lepton pairs, an additional factor $10^4$ (700 for $p_T>10$ GeV) suppression is required. However we can select the events with the different lepton flavours, such as� $e^+\mu^-$ and $e^-\mu^+$. This will eliminate the purely QED background. Unfortunately such $e^+\mu^-$ and $e^-\mu^+$ pairs could result from the $D\bar D$ (or $B\bar B$) decays. 
Most dangerous are the decays $D^+\to\pi^0\nu e^+$ 
(with branching ratio of $0.405\%$), $D^+\to\eta\nu e^+$  
(with branching ratio of $0.114\%$)
and $D_s\to\eta(\eta')\nu e^+$ 
(with combined branching ratio of $3.66\%$~\cite{PDG}); and the similar decays with the $e$ replaced by $\mu$.
That is we deal with the cross sections of the order of hundreds fb which must be suppressed more than 100 times.~\footnote{Analogous branching ratios for the $B$ meson decays are much smaller:
Br$(B^+\to\pi^0\nu l^+)=7.8\times 10^{-5}$, Br$(B^+\to\eta\nu l^+)=3.8\times 10^{-5}$ and Br$(B^+\to\eta'\nu l^+)=2.3\times 10^{-5}$~\cite{PDG}. Thus the $B$ mesons will not cause any serious problem. Moreover such events may be rejected since the vertex of the $B$� decay can be identified in the central� detector.}

Since the mass of the $D\bar D$ pair is much less than the mass of
the BSM objects that we are looking for, the corresponding suppression can be reached via the resolution of the forward proton detectors. The resolution must be sufficient to distinguish  the mass of $D\bar D$ ($B\bar B$) systems (of the order of 10-30 GeV) and the heavy object mass (larger than few hundreds GeV). Recall that actually the forward proton detectors (FpD) measure the proton momentum fractions $x^+$ and $x^-$; that is, they determine the mass $M^2_X=sx^+x^-$ and the rapidity $Y=0.5\ln(x^+/x^-)$ of the central system. Thus each individual FpD should provide a suppression of more than a factor of $\sqrt{100}=10$. This looks realistic. The only problem is to be sure that the signal was not washed out by proton dissociation.  
In other words we have to check that
\begin{itemize}
\item [(a)] There are no other high rapidity unobserved pions or photons which may carry away the corresponding energy. This can be done with the help of the ZDC calorimeters and Forward Shower Counters, as was described above. However, now we do not need the factor $10^6$ suppression in each outgoing proton direction, but instead a suppression of only about 10 times. This does not look at all pessimistic. Indeed, it may be done with the present detectors.

\item[ (b)] The resolution of the forward proton detectors must provide the possibility to distinguish an event with a central system  of mass, $M_C$, of about 
20 GeV from that of $M_C>200-500$ GeV, despite the fact that low-mass contribution is 100 times larger. Recall that we will also have to check the value of rapidity of the central system. That is, each forward proton detector 
should provide more then factor of 10 suppression. This also looks quite realistic.

\item [(c)]Final problem is the trigger. It might be difficult to organize the Level-1 trigger at high luminosities without the detection of large $E_T$ particles. However, as we said from the beginning, in our approach we have to work at relatively low luminosity with a mean number of interactions $\mu\sim 1$ per bunch crossing. In this case it may be sufficient to select low multiplicity events (in the central detector) with a lepton of $E_T>2-3$ GeV in coincidence with the two leading protons. 
\end{itemize}

It is worth mentioning that the BLM-based approach of \cite{BLM}
could open a way to test the models of dark matter, where
the mass splitting between the dark matter
particle and its charged (co-annihilation) partner
is so small that the co-annihilation partners become
long-lived (stable or meta-stable) at collider scales, see e.g. \cite{vvk}.
Then the final state to be observed is two forward protons plus
two anomalous charged tracks.

\section{Conclusion}
We have investigated the possibility of searching for a heavy invisible (BSM) object using the forward proton detectors; that is observing events with just forward protons and a large missing {\em longitudinal} energy. In such an experiment we have to be sure that the energy ``missed'' by the detectors was not carried away by undetected high rapidity SM particles. That is, to be sure that we observe `quasielastic' protons and not the decay products of some proton excitation (and dissociation $p\to M_X\to p+...$).
The probability of diffractive proton dissociation is very large ($\sim$ few mb) in comparison with the expected `invisible' signal of the order of fb.
  We have considered the characteristics of the `veto' detectors (like ZDC 
and FSC/ADA/ADC) that are required to suppress this background.
We argue that the only chance is to work at a relatively low luminosities where the mean number of pile-up events is $\mu\sim 1$ per bunch crossing.

Unfortunately, we have to conclude that for a completely invisible object, $X$, it will be {\em very} challenging to search for it in the exclusive $pp\to p+X+p$ reaction. The present ZDC, FSC, ADA, ADC detectors will not be able to sufficiently suppress the background.

The situation appears to be better for the `compressed mass' BSM
 scenarios where, after the decay of the new heavy object into 
a completely invisible particle (like the LSP in SUSY models) plus some
relatively low energy SM particles, these SM particles can be observed in the central detector. A good possibility is to observe in the central detector a pair of two different leptons (say $e^+\mu^-$) and nothing else. Then the required background suppression looks manageable.

\section{Acknowledgments}

We are grateful to Lucian Harland-Lang, Valya Khoze, Risto Orava and Albert De Roeck for 
useful discussions.
MGR thanks the IPPP at the University of Durham for hospitality.
 The work of MGR was supported by the RSCF grant 14-22-00281.
VAK thanks the Leverhulme Trust for an Emeritus Fellowship.

\end{document}